\documentclass[sigconf]{acmart}
\renewcommand\footnotetextcopyrightpermission[1]{}
\AtBeginDocument{%
  \providecommand\BibTeX{{%
    \normalfont B\kern-0.5em{\scshape i\kern-0.25em b}\kern-0.8em\TeX}}}

\copyrightyear{2025}
\acmYear{2025}

\copyrightyear{2025} 
\acmYear{2025} 
\makeatletter   
\def\@ACM@copyright@check@cc{} 
\makeatother 
\acmConference[]{}{}
\acmBooktitle{}\acmDOI{}
\acmISBN{}

\setcopyright{none}

\usepackage{xcolor}
\usepackage{caption}
\usepackage{subcaption}
\usepackage{multirow}
\usepackage{makecell}
\usepackage{array} 
\usepackage{longtable}
\usepackage{booktabs}
\usepackage{graphicx}
\usepackage{lscape}
\usepackage{ragged2e}
\usepackage{soul}
\usepackage{tablefootnote}
\usepackage{enumitem}
\usepackage{tabularx}
\usepackage{tikz}

\usepackage{tcolorbox}
\colorlet{LightYellow}{orange!30!}
\colorlet{LightRed}{red!20!}
\colorlet{LightBlue}{blue!20!}
\colorlet{LightViolet}{violet!20!}

\newcommand*{\circled}[2][]{\tikz[baseline=(C.base)]{
    \node[inner sep=0pt] (C) {\vphantom{1g}#2};
    \node[draw, circle, inner sep=1pt, yshift=0.7pt] 
        at (C.center) {\vphantom{1g}};}}

\colorlet{RED}{red}
\newcommand{\hr}[1]{{\textcolor{black}{
 #1
}}}

\newcolumntype{P}[1]{>{\raggedright\arraybackslash}p{#1}}

\begin{document}

\title[Impact Ambivalence]{Impact Ambivalence: How People with Eating Disorders Get Trapped in the Perpetual Cycle of Digital Food Content Engagement}
%


\author{Ryuhaerang Choi}
\affiliation{%
  \institution{KAIST}
  \city{Daejeon}
  \country{Republic of Korea}
}
\email{ryuhaerang.choi@kaist.ac.kr}

\author{Subin Park}
\affiliation{%
  \institution{KAIST}
  \city{Daejeon}
  \country{Republic of Korea}
}
\email{subin.park@kaist.ac.kr}

\author{Sujin Han}
\affiliation{%
  \institution{KAIST}
  \city{Daejeon}
  \country{Republic of Korea}
}
\email{sujinhan@kaist.ac.kr}

\author{Jennifer G. Kim}
\affiliation{%
  \institution{Georgia Institute of Technology}
  \city{Atlanta, Georgia}
  \country{USA}
}
\email{jennifer.kim@cc.gatech.edu}

\author{Sung-Ju Lee}
\affiliation{%
 \institution{KAIST}
 \city{Daejeon}
 \country{Republic of Korea}
}
\email{profsj@kaist.ac.kr}

\renewcommand{\shortauthors}{Ryuhaerang Choi et al.}



\begin{abstract}
Digital food content could impact viewers' dietary health, \hr{with individuals with eating disorders being particularly sensitive to it.} 
However, a comprehensive understanding of why and how these individuals interact with such content is lacking. To fill this void, we conducted \hr{exploratory (N=23) and in-depth studies (N=22)} with individuals with eating disorders to understand their motivations and practices of consuming digital food content. \hr{We reveal that participants engaged with digital food content for both disorder‑driven and recovery‑supporting motivations, leading to conflicting outcomes. This \emph{impact ambivalence}, the coexistence of recovery‑supporting benefits and disorder‑exacerbating risks, sustained a cycle of quitting, prompted by awareness of harm, and returning, motivated by anticipated benefits. We interpret these dynamics within dual systems theory and highlight how recognizing such ambivalence can inform the design of interventions that foster healthier digital food content engagement and mitigate post-engagement harmful effects.} 
\end{abstract}

\maketitle

\section{Introduction}

\textbf{Caution}: \emph{This paper discusses eating disorders and includes media that could potentially be a trigger to those dealing with eating disorders. Please use discretion when reading and disseminating this paper.}

Digital media's exponential growth has made food content ubiquitous and diversified. On social media, the popularity of food videos (e.g.,~eating and cooking broadcasts) and images has significantly increased~\cite{kang2020popularity, mejova2016fetishizing, de2005tv, adema2000vicarious}. On TikTok, the hashtag \#TikTokFood has been \hr{posted over 5M~times~\cite{tiktokfood}} and viewed over 43~billion times~\cite{tiktokwrapped2021}, and Instagram has over \hr{313~million} posts tagged with \#foodporn~\cite{InstaFoodporn}. 
\hr{The online food market---especially app-based delivery that trades on visually enticing food imagery---has kept surging: delivery accounted for 21\% of global consumer food service spending in 2024, up from 9\% in 2019~\cite{foodservice_delivery}.} 

A growing body of research has examined the impact of increased exposure to various digital food content on eating habits and dietary health~\cite{petit2016can, kinard2016insta, ofli2017saki, vermeir2020visual, pope2015viewers, choi2024foodcensor, wu2024digging, wang2024viral}. 
The trend of watching and posting food images and videos with catchy hashtags has been associated with inadequate eating habits~\cite{yan2019digital, qutteina2019media}, unhealthy food choices~\cite{peng2018feast, basso2018taste}, and overeating~\cite{bodenlos2013watching, halford2004effect} through various visual and social factors~\cite{seal2022can}. Browsing and ordering food using online shopping and food delivery services are also known to involve unhealthy food choices~\cite{yeo2017consumer}, unhealthy eating~\cite{eu2021consumers}, and obesity~\cite{stephens2020food}. Moreover, many digital food content features take-away food~\cite{kang2020popularity}, and combining eating content with online food delivery services could lead to unhealthy lifestyles. 

Recent studies have highlighted that \emph{individuals with eating disorders} (EDs) engage with digital food content significantly more than the general population~\cite{bachner2018lives} and are particularly sensitive to such content~\cite{jin2018interactive, legenbauer2004anticipatory}. 
When coupled with frequent unconscious exposure to digital food content, it could exacerbate ED symptoms, such as binge eating and obsessive food thoughts~\cite{choi2024foodcensor, kinkel2022food}. 
Despite these findings, there remains a significant gap in understanding why individuals with EDs engage with digital food content and how such engagement ultimately contributes to their disordered eating in their daily lives. In this regard, we aim to shed light on the experience of people with disordered eating behaviors with digital food content by answering the following question: \textit{Why and how do people with EDs interact with digital food content?} 

To answer this question, we conducted two rounds of studies with individuals with eating disorders:~one round of an exploratory study (N=23) and another round of an in-depth study (N=22). 
The exploratory study assesses how likely individuals with EDs are to be concerned about associations between digital food content and disordered eating, and identifies which digital food content types are associated with disordered eating. 
Notably, we refrained from bringing up digital food content, allowing participants to broach the subject voluntarily. Sixteen participants out of twenty-three voluntarily linked digital food content to their ED symptoms. Our analysis identified two representative digital food content positively associated with ED symptoms: food media~(e.g.,~images and videos) and food delivery apps. \hr{In particular, we found that food media played an \emph{ambivalent} role; sometimes used for managing eating-related urges, yet at other times acting as a trigger for binge eating. }


\hr{To further understand the underlying dynamics of this ambivalence, we conducted in-depth interviews focusing on ED individuals' motivations and practices of consuming digital food content.} Going beyond prior research's emphasis on the detrimental impacts of digital food content on ED populations~\cite{kinkel2022food, qutteina2019media, yan2019digital}, \hr{our in-depth study examined how this ambivalence permeates participants' engagement from motivations that draw them to such content, to the impacts it produces, and the cycles of avoidance and relapse it sustains.} 

\hr{This tension echoes the concept of \emph{attitudinal ambivalence}, the coexistence of positive and negative evaluations toward the same object or behavior~\cite{thompson2014let, van2009agony, schneider2017mixed}. In our study, we first observed the coexistence of recovery-supporting and disorder-exacerbating impacts within the same digital food content engagement. We term this coexistence \emph{impact ambivalence}, extending attitudinal ambivalence to the consequence domain of behavior. Importantly, the recognition of such conflicting impacts shapes individuals' anticipations of future engagements, which then feed into \emph{attitudinal ambivalence}: they come to both desire engagement, expecting benefits, and resist it, anticipating harms. Such impact ambivalence not only explains individuals' recurring cycle of quitting and returning to digital food content, but also surfaces design opportunities often overlooked by binary harm-reduction perspectives. Recognizing this ambivalence enables the identification of moments where supportive and harmful effects intersect, informing interventions that guide users' decisions before engagement and mitigate adverse impacts afterward.
}

The main contributions of this paper are summarized as follows. 
\begin{itemize}
    \item We reveal the perceptions, motivations, impacts, and engagement patterns with various digital food content among individuals with eating disorders. 
    Extending beyond prior literature, which primarily focuses on its detrimental impacts on individuals with EDs, our findings provide deeper insights into why and how these individuals engage with such content, highlighting its overlooked recovery-supporting potential. 
    \item We articulate the cognitive mechanisms underlying how individuals with EDs interact with digital food content by integrating the concept of \emph{impact ambivalence} into dual systems theory (Section~\ref{sec:discussion:ambivalence}). While the double-sided impacts of digital food content in the general population indicate distinct positive (e.g.,~entertainment) and negative (e.g.,~triggering overeating) outcomes, \emph{impact ambivalence} reveals the coexistence of conflicting outcomes, \hr{specifically} recovery-supporting benefits and ED-exacerbating risks. This perspective offers deeper insights into the unique challenges faced by individuals with EDs.
    \item Drawing on a thorough understanding of the cognitive mechanisms behind their content engagement, we propose intervention designs that promote informed and mindful interactions with digital food content \hr{and mitigate} its effects after engagement.
\end{itemize}

Our study complements prior research on digital food content by \hr{foregrounding the experiences of individuals with eating disorders, providing unique insights into their engagement patterns, and informing the design of digital interventions that better reflect their needs and struggles.}
\section{Related Work}

\subsection{Eating Disorders and Digital Content Use}

The intersection of digital content use and eating disorders has been a focal point in the HCI community~\cite{feuston2020conformity, pater2021charting, pater2019notjustgirls}. Studies have identified that extensive engagement with digital media, including pro-eating disorder content and imagery showcasing idealized body images, correlates with the onset and exacerbation of eating disorders~\cite{chancellor2017multimodal, gak2022distressing}. Research has also underpinned the critical role of digital environments, which often emphasize thinness and perfection as desirable body standards, in shaping individuals' perceptions of body image and eating behaviors~\cite{aparicio2019social}. In addition, such social media environments encouraged individuals to constantly compare individuals with influencers and celebrities, leading to negative body image and low self-esteem, driving individuals to engage in disordered eating behaviors to meet unrealistic ideals~\cite{pedalino2022instagram}. Individuals vulnerable to body dissatisfaction and disordered eating behaviors are particularly susceptible to these media messages, as they fuel the internalization of harmful beauty ideals~\cite{murraystuart2018contribution}. The consistent engagement in such content contributed to the persistence and intensification of eating disorder symptoms.



The interplay between digital content and eating disorders has underscored the need for a comprehensive investigation that considers the multifaceted nature of the impact of digital content on eating disorders~\cite{chancellor2016post, chancellor2016thyghgapp}. While a substantial body of research has explored the link between various digital content consumption and eating disorders, there is a notable gap in our understanding of the specific association between food-related content, which is closely connected to dietary health~\cite{petit2016can, kinard2016insta, ofli2017saki, vermeir2020visual}, and disordered eating behaviors.

\subsection{An Overarching Presence of Digital Food Content}

Digital food content comprises diverse multimedia materials that revolve around the food theme. This encompasses various content types, including captivating recipe videos, tantalizing food imagery, informative cooking tutorials, and engaging food blogs. Moreover, the proliferation of food delivery apps that present visually captivating food pictures has expanded digital food content, providing users with seamless access to various culinary offerings through their digital devices~\cite{maimaiti2018we}.

In today's digital age, the prevalence of digital food content is palpable across various online platforms~\cite{kang2020popularity, mejova2016fetishizing, weber2021s, tiktokwrapped2021}. Individuals constantly engage in various food-related content, from social media to video-sharing platforms, dedicated food websites, and mobile applications. This content not only caters to culinary enthusiasts seeking inspiration and knowledge~\cite{de2005tv, wang2022tiktok} but also to a broader audience drawn by the visual allure and cultural narratives associated with food~\cite{spence2016eating, adema2000vicarious, peng2018feast}. As a result, the pervasive presence of digital food content has transformed how people engage with and perceive food in their daily lives~\cite{yan2019digital}.

This omnipresence of digital food content underscores the need to explore its impact on eating behaviors, especially among individuals with eating disorders who are especially susceptible to such content~\cite{jin2018interactive, legenbauer2004anticipatory}. The subsequent section delves into the intricate dynamics of digital food content's influence on individuals' eating behaviors, shedding light on its constructive and detrimental effects.

\subsection{Double-sided Impact of Digital Food Content on Eating Behaviors}

Digital food content exerts a complex influence on individuals' eating behaviors, presenting a dual-sided impact characterized by both constructive and detrimental facets~\cite{strand2020mukbang, wang2024viral}. While many individuals engage with digital food content for inspiration, recipe ideas, and cultural connections to food~\cite{de2005tv}, the comprehensive impact of such media on individuals with eating disorders, who are much more sensitive to digital food content than the general population~\cite{schienle2009binge, lee2017impaired}, has received limited attention. The relationship between individuals' consumption of digital food content, including the engagement with eating broadcasts (\textit{Mukbang}), and the potential escalation of disordered eating behaviors has garnered recognition in prior research, primarily focusing on the general population~\cite{von2023associations, anari2023relationship}. However, these investigations have yielded varying insights and have yet to explicitly target individuals with eating disorders.

Certain instances of digital food content consumption have been associated with fostering positive eating habits and healthier dietary choices~\cite{drivas2024whatieatinaday}. Informative cooking tutorials, recipe demonstrations, and nutrition education videos can empower individuals to experiment with wholesome meal options and develop improved culinary skills~\cite{wang2022tiktok, ventura2021socialfood}. These platforms potentially serve as educational tools, equipping users with knowledge to make informed decisions about their food intake. Such exposure can lead to a sense of agency in managing dietary preferences and restrictions, encouraging healthier relationships with food~\cite{donnar2017food}. Furthermore, for certain individuals, engaging with digital food media provides a sense of presence that helps alleviate their loneliness~\cite{anjani2020people}. 
However, it is essential to acknowledge that these potential positive effects are not uniformly experienced, and their effectiveness can depend on individual and personal contexts~\cite{strand2020mukbang, seal2022can}.

Conversely, a growing body of research has highlighted the potentially detrimental impact of digital food content, especially when exacerbating unhealthy eating habits~\cite{pope2015viewers, spence2016eating}. The omnipresence of visually appealing, calorie-laden food imagery in digital content can escalate disordered eating behaviors~\cite{bodenlos2013watching, ventura2021socialfood}. Studies have pointed out that exposure to images and videos of excessive food consumption, particularly those featuring eating challenges or binge eating, can inadvertently trigger or reinforce unhealthy eating patterns~\cite{spence2019digital}. Such content inadvertently glorifies or normalizes overconsumption~\cite{kircaburun2021problematic}. Additionally, the visual appeal of enticing food displays and the sensory experience of hearing eating sounds may inadvertently intensify cravings and compulsive eating tendencies~\cite{spence2016eating, anjani2020people}. Recent research proposed censoring food visuals to mitigate these adverse effects and demonstrated that preventing unconscious consumption of digital food content can help manage eating disorder conditions in daily life~\cite{choi2024foodcensor}.

In addition, food delivery apps, despite their convenience, have been implicated in negatively influencing eating habits~\cite{portingale2023tonight}. These platforms often feature visually appealing images of indulgent and calorie-rich foods strategically designed to entice users~\cite{kapoor2018technology}. Such visual appeals can lead to impulsive and unplanned consumption online~\cite{liu2013website}. Moreover, frequent use of food delivery apps can normalize excessive consumption and encourage a sedentary lifestyle, further exacerbating the risk of weight gain and related health issues~\cite{stephens2020food, maimaiti2018we}. The targeted marketing tactics employed by these apps might amplify the temptation to order food impulsively, making it challenging for individuals, especially those susceptible to disordered eating tendencies, to maintain a balanced and mindful approach to their eating habits.

The current research landscape lacks a profound understanding of why and how individuals with eating disorders engage with digital food content—a realm encompassing spontaneous interactions and passive influences. Existing studies are confined to quantitative survey-based methods, such as examining the correlation between the presence of eating disorders and the frequency of viewing eating broadcast~\cite{kircaburun2021problematic, kircaburun2021compensatory, von2023associations, kim2021binge, anari2023relationship, bachner2018lives}. While these studies are meaningful in establishing links between eating disorders and digital food content, they leave the deeper psychological and behavioral dynamics largely unexplored, creating a significant gap in our understanding. 
These studies suggested further research into the various aspects of digital food content and media types that potentially promote disordered eating~\cite{kircaburun2021compensatory, kim2021binge}. 
The goal of our research is to shed light on these intricate dynamics and unravel the profound impact these interactions have on individuals' attitudes, behaviors, and overall well-being. We aim to provide insights into the mechanisms underpinning the sway of digital food content on this population, paving the way for effective and tailored interventions and support mechanisms on digital food content platforms.

\section{Exploratory Study} 
\label{sec:exploratoryStudy}

Before delving deeply into the investigation of how digital food content affects individuals with EDs, we conducted an IRB-approved exploratory study to broadly assess the extent to which they connect with digital food content concerning their ED symptoms. 

\subsection{Participants}

\begin{table*}[ht]
\resizebox{\textwidth}{!}{
\begin{tabular}{rrccccrccl}
\Xhline{2\arrayrulewidth}
\multicolumn{1}{c}{\textbf{P}} & \multicolumn{1}{c}{\textbf{Age}} & \textbf{Gender} & \textbf{\begin{tabular}[c]{@{}c@{}}ED\\ Population\end{tabular}} & \textbf{\begin{tabular}[c]{@{}c@{}}Diagnosis\\ (formal/self)\end{tabular}} & \textbf{\begin{tabular}[c]{@{}c@{}}ED\\ Duration\end{tabular}} & \multicolumn{1}{c}{\textbf{\begin{tabular}[c]{@{}c@{}}EDE-Q\\ Score\end{tabular}}} & \textbf{Interview}   & \textbf{Survey}      & \multicolumn{1}{c}{\textbf{Stated Digital Food Content}} \\ \hline\hline
1                              & 18                               & Female          & BN                                                               & formal                           & 1 years 2 months                                               & 4.83                                         & \multicolumn{1}{l}{} & Yes                  &                                                          \\
2                              & 19                               & Female          & BED                                                              & self                             & 4 years                                                        & -                                            & \multicolumn{1}{l}{} & Yes                  &                                                          \\
3                              & 20                               & Female          & BN                                                               & formal                           & 5 years                                                        & 5.39                                         & Yes                  & \multicolumn{1}{l}{} & food content, food delivery apps                         \\
4                              & 20                               & Female          & BED                                                              & self                             & 7 years                                                        & 5.48                                         & Yes                  & Yes                  & food content                                             \\
5                              & 21                               & Female          & BN                                                               & self                             & 1 year                                                         & -                                            & \multicolumn{1}{l}{} & Yes                  &                                                          \\
6                              & 23                               & Female          & BN                                                               & formal                           & 4 years                                               & 4.68                                         & Yes                  & Yes                  & food delivery apps                                       \\
7                              & 23                               & Female          & BN                                                               & self                             & 4 years                                                        & 4.22                                         & Yes                  & Yes                  & food content                                             \\
8                              & 23                               & Female          & BN                                                               & formal                           & 6 years 5 months                                               & 4.98                                         & Yes                  & Yes                  & food delivery apps                                       \\
9                              & 24                               & Not to disclose & BN                                                               & self                             & 1 year                                                         & 4.32                                         & Yes                  & \multicolumn{1}{l}{} & food delivery apps                                       \\
10                             & 24                               & Female          & BN                                                               & formal                           & 4 years                                                        & 4.55                                         & Yes                  & Yes                  & food content, food delivery apps                         \\
11                             & 24                               & Female          & BED                                                              & formal                           & 5 years                                                        & 3.97                                         & Yes                  & Yes                  & food content, food delivery apps                         \\
12                             & 24                               & Female          & BED                                                              & self                             & 9 years                                                        & -                                            & \multicolumn{1}{l}{} & Yes                  & food delivery apps                                       \\
13                             & 25                               & Female          & BN                                                               & formal                           & 2 years 2 months                                               & 5.03                                         & Yes                  & \multicolumn{1}{l}{} & food content                                             \\
14                             & 25                               & Non-binary      & BN                                                               & self                             & 3 years                                                        & 4.16                                         & Yes                  & \multicolumn{1}{l}{} & food content                                             \\
15                             & 26                               & Female          & BN                                                               & self                             & 3 years                                                        & 3.52                                         & Yes                  & Yes                  & food content, food delivery apps                         \\
16                             & 27                               & Female          & BN                                                               & formal                           & 13 years                                                       & -                                            & \multicolumn{1}{l}{} & Yes                  &                                                          \\
17                             & 29                               & Male            & BED                                                              & self                             & 3 years                                                        & -                                            & \multicolumn{1}{l}{} & Yes                  & food content                                             \\
18                             & 30                               & Female          & BED                                                              & self                             & 2 months                                                       & -                                            & \multicolumn{1}{l}{} & Yes                  & food content                                             \\
19                             & 30                               & Not to disclose & BN                                                               & formal                           & 4 years                                                        & 3.98                                         & Yes                  & Yes                  & food delivery apps                                       \\
20                             & 30                               & Female          & BN                                                               & formal                           & 15 years                                                       & -                                            & \multicolumn{1}{l}{} & Yes                  &                                                          \\
21                             & 36                               & Female          & BN                                                               & formal                           & 17 years                                                       & 4.66                                         & Yes                  & \multicolumn{1}{l}{} &                                                          \\
22                             & 38                               & Female          & BED                                                              & self                             & 6 years 11 months                                              & -                                            & \multicolumn{1}{l}{} & Yes                  &                                                          \\
23                             & 43                               & Female          & BED                                                              & self                             & 20 years                                                       & -                                            & \multicolumn{1}{l}{} & Yes                  & food content                                            
                    \\ \Xhline{2\arrayrulewidth}
\end{tabular}
}
\caption{Exploratory study participants' demographic, eating disorder information, and stated digital food content. The listed items (i.e.,~food content and food delivery apps) in the stated digital food content are identified from open-ended responses in surveys and interviews. BED refers to Binge Eating Disorder and BN stands for Bulimia Nervosa. BED is characterized by recurrent episodes of excessive eating in a short timeframe~(e.g.,~2~hours) and feeling a loss of control over their eating behavior~\cite{dingemans2002binge}. BN involves a cycle of binge eating and compensatory behaviors, such as self-induced vomiting~\cite{fairburn1986diagnosis}. 
}
\label{table:exploratory:participants}
\vspace{-0.5cm}
\end{table*}

We recruited 23~participants (aged 18-43, mean=26.2~years; 19~females, 1~male, 1~non-binary, 2~chose not to disclose) through an advertisement post on an online community for people with EDs under the permission of the moderators in South Korea~\cite{KoreanEatingDisorderSocialSupportCommunity}. The participant gender composition is predominantly female, reflecting universal ED demographics (87.4\%~of identified South Korean binge eating disorder patients being female~\cite{KoreaBEDProportion}, and threefold higher lifetime prevalence of binge eating disorder and bulimia nervosa in females in the US~\cite{hudson2007prevalence, galmiche2019prevalence}). 
Table~\ref{table:exploratory:participants} provides an overview of participant demographics, ED information, and types of digital food content mentioned by each participant. 
Our study intentionally included individuals who were either formally diagnosed or self-diagnosed, as has been done in prior research~\cite{choi2024foodcensor, feuston2020conformity, choi2025private}. By doing so, we aimed to encompass a population that may not have sought professional treatment, providing insights into the often overlooked experiences of these individuals within a clinical context. 

\subsection{Study Protocol}

The study involved an online survey to investigate how people with EDs associate digital food content with their ED symptoms. Additionally, we conducted remote semi-structured interviews to gain a deeper understanding. We employed the Eating Disorder Examination Questionnaire 6.0 (EDE-Q) to assess if our participants represented the eating disorder population~\cite{fairburn2008eating}.\footnote{Higher scores on the EDE-Q indicate problematic eating behaviors and attitudes. The EDE-Q scores of our study's interviewees (4.56±0.56) were generally higher than those of individuals diagnosed with ED~(4.02±1.28)~\cite{aardoom2012norms}.} 
To address the social stigma associated with EDs~\cite{puhl2015stigma}, participants were given the option to opt out of any part of the study, which consists of a survey, an interview, and the EDE-Q, they find uncomfortable. 

In surveys, we intentionally refrained from introducing digital food content to accurately assess how participants perceived its influence on their symptoms. 
Instead, we asked about their computing device usage associated with disordered eating compared with normal meals. Specifically, survey questions included are: \textit{Are there any differences in your digital device usage before/during/after eating disorder episodes than without eating disorder episodes? If so, which device(s) do you use, and for what purpose?} This approach allowed us to identify unique device usages, including consumption of digital food content, associated with their disordered eating. 

Each Zoom interview, lasting approximately thirty minutes to one hour, was recorded and transcribed after the interview with consent. Given the stigma surrounding EDs~\cite{puhl2015stigma}, we allowed the participants to freely turn off their cameras during the interview. In the interviews, participants shared their experiences regarding how various digital device usages are associated with and result in disordered eating. We also inquired about any apps and content they perceived to affect their ED symptoms. For participants who did not respond to the survey, we also asked the survey questions during the interview. We did not bring up digital food content unless participants voluntarily mentioned it, which is consistent with our survey approach. Only the interviewees, not survey-only respondents, were rewarded approximately USD~23 for a one-hour interview. \hr{The study materials, including survey questions, interview questions and participants' responses, were administered in Korean and subsequently translated into English. All translations were reviewed by the authors to ensure accuracy and preserve the integrity of participants' original statements.}

\subsection{Analysis}

We followed the inductive thematic analysis~\cite{braun2012thematic}. The first author initiated the data familiarization process by transcribing the interviews and used the inductive approach to open-code the survey responses and interview transcripts. To comprehensively understand and validate the initial code, every author reviewed the data and emerging themes throughout regular discussions. Then, we developed two themes that capture digital food content types associated with their ED symptoms. Finally, we titled each theme and gave a description. Note that we report only themes that involve digital food content as other types of digital content, such as body images on social media, have been disclosed from previous studies that investigated triggering multimedia content for ED patients~\cite{feuston2020conformity, gak2022distressing}.

\subsection{Results}

\hr{Sixteen out of twenty-three participants voluntarily linked digital food content to their disordered eating behaviors, revealing a nuanced relationship. Our analysis identified two primary types of content associated with such behaviors: food media and food delivery apps. Specifically, eleven participants mentioned food media, including recipe blogs, food photos, chewing or crunching ASMR videos,\footnote{An Autonomous Sensory Meridian Response (ASMR) refers to a tingling sensation in response to specific visual, auditory, or tactile stimuli. ASMR videos often feature soothing voices, tapping, eating, and other soothing sounds.} and eating broadcasts. Nine participants referred to food delivery apps. These findings reaffirm the connection between digital food content and unhealthy dietary habits~\cite {bodenlos2013watching, spence2019digital}.}

\hr{Participants' narratives revealed that food media played an ambivalent role in their experiences with disordered eating, serving both as a self-regulation strategy and a trigger for binge episodes. Several participants described watching food-related videos in an effort to suppress the urge to binge eat or to gain emotional comfort. For example, P15 shared, ``\textit{I watched Mukbangs to try to stop myself from binge eating to get some vicarious satisfaction. I watch them to suppress the urge.}'' She also noted that she ``\textit{sometimes watches videos of others who binge and regret it, especially when feeling stressed. It gives me a sense of solidarity and reminds me that I'm not alone, and that recovery is still possible by seeing people in such videos who make up their minds again even after such relapses.}'' Likewise, P3 shared, ``\textit{I've tried watching slow eating videos while eating, to help myself pace and prevent binge episodes.}''}

\hr{Yet, the same form of media often triggered harmful outcomes as emphasized in the previous study~\cite{choi2024foodcensor}. Some participants, including P3 and P15, who described food media as a helpful coping tool, also reported that it sometimes led to binge eating. P7 explained, ``\textit{When I repeatedly view such things (suggested food media), I unconsciously feel the urge to eat it. It would be okay if I could eat it without feeling guilty about it, but if I, even once, feel guilty (because it would lead to weight gain), it piles up on and on, pushing me to binge eat. So, in the long run, it does have an impact.}''}

\hr{Through our exploratory study, we identified the \emph{ambivalent impact} of food media---both supporting and undermining recovery efforts within a single engagement. To further understand the underlying dynamics of this ambivalence, we conducted a follow-up in-depth interview study focusing on (1)~individuals' motivations for engaging with digital food content, (2)~the conflicting impacts such content has on their eating disorder experiences, and (3)~the strategies they adopt to avoid exposure to triggering content. This deeper investigation aims to inform the design of digital interventions that acknowledge both the harms and the personally meaningful benefits embedded in everyday interactions with food content among individuals with eating disorders.}

\section{In-depth Study}

\subsection{Participants}

\begin{table*}[ht]
\resizebox{\textwidth}{!}{
\begin{tabular}{rrlcccrl}
\Xhline{2\arrayrulewidth}
\multicolumn{1}{c}{\textbf{P}} & \multicolumn{1}{c}{\textbf{Age}} & \multicolumn{1}{c}{\textbf{Gender}} & \multicolumn{1}{c}{\textbf{\begin{tabular}[c]{@{}c@{}}ED\\ Population\end{tabular}}} & \textbf{\begin{tabular}[c]{@{}c@{}}Diagnosis\\ (formal/self)\end{tabular}} & \textbf{\begin{tabular}[c]{@{}c@{}}ED\\ Duration\end{tabular}} & \multicolumn{1}{c}{\textbf{\begin{tabular}[c]{@{}c@{}}EDE-Q \\ Score\end{tabular}}} & \multicolumn{1}{c}{\textbf{Stated Digital Food Content}} \\ \hline\hline
1                              & 18                               & Female                              & BN                                                                                   & self                                                                       & 1 year                                                         & 4.86                                                                                & food media, food delivery apps                          \\
2                              & 20                               & Female                              & BN                                                                                   & self                                                                       & 3 months                                                       & 3.84                                                                                & food media                                               \\
3                              & 20                               & Female                              & BN                                                                                   & self                                                                       & 1 year                                                         & 4.67                                                                                & food media, food delivery apps                           \\
4                              & 21                               & Female                              & BN                                                                                   & formal                                                                     & 2 years                                                        & 4.38                                                                                & food media                                               \\
5                              & 22                               & Female                              & BED                                                                                  & self                                                                       & 3 year                                                         & 4.41                                                                                & food media                                               \\
6                              & 22                               & Female                              & BN                                                                                   & formal                                                                     & 3 years 1 month                                                & 5.30                                                                                & food media                                               \\
7                              & 22                               & Female                              & BN                                                                                   & formal                                                                     & 10 years                                                       & 5.11                                                                                & food media                                               \\
8                              & 23                               & Female                              & BN                                                                                   & self                                                                       & 3 years                                                        & 2.72                                                                                & food media, food delivery apps                           \\
9                              & 23                               & Female                              & BN                                                                                   & formal                                                                     & 4 years                                                        & 4.59                                                                                & food media, food delivery apps                           \\
10                             & 24                               & Female                              & BED                                                                                  & self                                                                       & 2 years                                                        & 4.19                                                                                & food media                                               \\
11                             & 24                               & Female                              & BN                                                                                   & formal                                                                     & 6 year 5 months                                                        & 5.27                                                                                & food media                                               \\
12                             & 25                               & Female                              & BN                                                                                   & self                                                                       & 5 years                                                        & 4.82                                                                                & food media, food delivery apps                           \\
13                             & 26                               & Female                              & BN                                                                                   & self                                                                       & 2 years                                                        & 4.71                                                                                & food media                                               \\
14                             & 26                               & Female                              & BED                                                                                  & self                                                                       & 2 years                                                        & 5.36                                                                                & food media, food delivery apps                           \\
15                             & 29                               & Female                              & BN                                                                                   & self                                                                       & 1 year                                                         & 4.96                                                                                & food media                                               \\
16                             & 29                               & Female                              & BED                                                                                  & self                                                                       & 2 years                                                        & 4.47                                                                                & food media, food delivery apps                           \\
17                             & 33                               & Female                              & BN                                                                                   & formal                                                                     & 7 years 2 months                                               & 5.21                                                                                & food media, food delivery apps                           \\
18                             & 33                               & Female                              & BN                                                                                   & self                                                                       & 15 years                                                       & 5.03                                                                                & food media, food delivery apps                           \\
19                             & 34                               & Female                              & BED                                                                                  & formal                                                                     & 10 years                                                       & 5.03                                                                                & food media, food delivery apps                           \\
20                             & 34                               & Female                              & BN                                                                                   & formal                                                                     & 15 years                                                       & 4.32                                                                                & food delivery apps                                       \\
21                             & 39                               & Male                                & BED                                                                                  & self                                                                       & 10 years                                                       & 2.74                                                                                & food media                                               \\
22                             & 41                               & Female                              & BN                                                                                   & self                                                                       & 20 years                                                       & 0.72                                                                                & food delivery apps \\  \Xhline{2\arrayrulewidth}
\end{tabular}}
\caption{In-depth study participants' demographics, ED information, and stated digital food content. The stated digital food content items are identified from open-ended interview responses. 
}
\label{table:in-depthstudy:participants_ede}
\vspace*{-0.3cm}
\end{table*}

We recruited 22~participants (aged~18-41, mean=26.7~years; 21~females and 1~male) through advertisement posts on the online social support communities for people with EDs under the permission of the moderators in South Korea~\cite{KoreanEatingDisorderSocialSupportCommunity, KakaoSocialSupportChatroom}. 
Table~\ref{table:in-depthstudy:participants_ede} presents participants' demographics and ED diagnosis information. 
Two in-depth study participants, P9 and P11 (P6 and P8 in the exploratory study, respectively), also participated in our exploratory interviews. 
To be eligible for the study, participants must (1)~be over 18 years old and (2)~have an ED. Since many people do not seek treatment for their EDs, eligibility for this study, as in our exploratory study, was not contingent upon a diagnosis. Participants, however, had to identify themselves as having ED symptoms (e.g.,~restriction of food intake, binge eating, or compensatory behaviors). Again, by recruiting users who did not seek treatment or had discontinued it, we could shed light on the needs of users whose struggles are largely ignored in a clinical setting.

\subsection{Study Protocol}

All phases of our study were IRB-approved and conducted remotely due to the high social stigma of EDs~\cite{puhl2015stigma}. Participants responded to a survey via email before the interview. The survey included the EDE-Q and questions about their demographics and ED~symptoms.\footnote{The average EDE-Q score of the participants was 4.40±1.09, compared with averages of individuals diagnosed with ED~(4.02±1.28)~\cite{aardoom2012norms}.} 
Following this, we conducted semi-structured interviews with all participants on Zoom, each lasting approximately thirty minutes to one hour. We allowed participants to turn off their cameras. Every interview was recorded and transcribed with consent from the participants. Our interview delved into the motivations and practices of digital food content consumption by asking questions about participants’ experiences. We encouraged participants to share their varied experiences with different types of food media, covering various food videos, pictures, and posts. The detailed interview protocol is in our Supplementary Materials. \hr{The study materials, including interview questions and participants' responses, were administered in Korean and subsequently translated into English. All translations were reviewed by the authors to ensure accuracy and preserve the integrity of participants' original statements.}

\subsection{Analysis}

We conducted the inductive thematic analysis~\cite{braun2012thematic} of the interview responses to investigate why and how individuals with EDs interact with digital food content and how such content relates to disordered eating behaviors. After transcribing the interviews, the first and second authors individually looked through all transcriptions to fully understand the interview content. They independently generated initial codes and, through iterative discussions on emerging themes, addressed inconsistencies, resolved disagreements, and ensured that the data strongly supported the identified themes. To finalize the analysis, all authors participated in comprehensive discussions. Once the analysis was completed, each theme was titled and described.

\section{In-Depth Study Results}
\label{sec:results}


\hr{Building on our exploratory study's observation of impact ambivalence---the coexistence of recovery-supporting and ED-exacerbating effects within the same engagement---we structured our in-depth analysis to examine how this ambivalence manifests across participants' motivations, perceived impacts, and avoidance practices.}

\subsection{Patterns of Digital Food Content Consumption}
\label{res:patterns}

\begin{figure*}
    \centering
    \includegraphics[width=\textwidth]{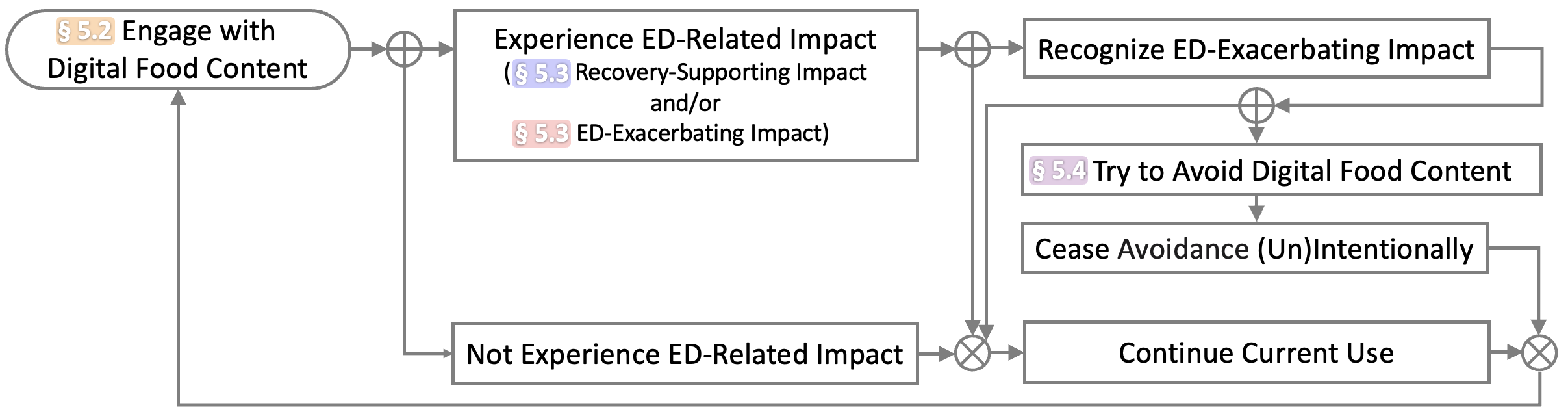}
    \vspace{-0.3cm}
    \caption{A flowchart describing interaction patterns with digital food content. The OR symbol ($\oplus$) denotes that the process continues in multiple branches. The Summoning Junction symbol ($\otimes$) indicates that multiple branches converge into a single process. 
    }
    \Description{Figure 1 is the flowchart that depicts patterns of digital food content engagement. The detailed explanation of these patterns is described in Section 5.1.}
    \label{fig:patterns_dfc}
\end{figure*}

\hr{Before delving into specific motivations, impacts, and avoidance practices, we outline the overarching patterns of participants’ engagement with digital food content. This overview provides context for how these different aspects unfold over time and interact with one another (Figure~\ref{fig:patterns_dfc}).}

\hr{Across the interviews, participants described their engagement as a recurring sequence:}
\begin{enumerate}
    \item \hr{\textbf{Engage with Digital Food Content:} Participants began consuming digital food content, often by chance or with specific motivations, such as seeking comfort, managing urges, or browsing out of habit.}
    \item \hr{\textbf{Recovery‑Supporting Impact:} Early in their engagement, they felt the content satisfied their motivations and did not recognize harmful effects.}
    \item \hr{\textbf{Recognize ED‑Exacerbating Impact:} Over time, many noticed that the same content often worsened their ED symptoms.}
    \item \hr{\textbf{Try to Avoid Digital Food Content:} This realization led participants to take deliberate steps to limit or stop exposure, such as changing settings, unfollowing channels, or uninstalling apps.}
    \item \hr{\textbf{Cease Avoidance (Un)Intentionally:} Despite avoidance efforts, participants frequently resumed engagement, due to loss of self‑control, incidental exposure, or expectations of gaining benefits without harm, re‑entering the cycle.}
\end{enumerate}

\hr{We detail each stage in the following sections, examining the underlying motivations (Section~\ref{res:motivation}), the ambivalent impacts experienced (Section~\ref{res:dfc}), and the strategies and challenges of avoidance (Section~\ref{res:efforts}).}

\subsection{\hr{Recovery-Driven and Disorder-Driven} Motivations for Digital Food Content Consumption} 
\label{res:motivation}

\begin{table*}[t]
\resizebox{\textwidth}{!}{
\begin{tabular}{lll}
\Xhline{2\arrayrulewidth}
\textbf{Digital Food Content}       & \textbf{\tcbox[on line, frame empty, boxsep=0pt,left=2pt,right=2pt,top=2pt,bottom=2pt,colback=LightYellow]{Motivation}}                             & \textbf{Stated Subcategories of Digital Food Content}                                                    \\ \hline \hline
\multirow{4}{*}{Food Media}         & \tcbox[on line, frame empty, boxsep=0pt,left=2pt,right=2pt,top=2pt,bottom=2pt,colback=LightYellow]{M1} Expectations for Vicarious Satisfaction        & \begin{tabular}[c]{@{}l@{}}eating broadcast, tasting video, \\ new food products showcasing video/post\end{tabular} \\ \cline{2-3} 
                                    & \tcbox[on line, frame empty, boxsep=0pt,left=2pt,right=2pt,top=2pt,bottom=2pt,colback=LightYellow]{M2} Expectations for Comfort from Shared Experience & binge eating broadcast/post                                                                                   \\ \cline{2-3} 
                                    & \tcbox[on line, frame empty, boxsep=0pt,left=2pt,right=2pt,top=2pt,bottom=2pt,colback=LightYellow]{M3} Seeking Motivation for Change                          & binge eating broadcast                                                                                            \\ \cline{2-3} 
                                    & \tcbox[on line, frame empty, boxsep=0pt,left=2pt,right=2pt,top=2pt,bottom=2pt,colback=LightYellow]{M4} Expectations for Relief by Cooking along with the Media & cooking video                                                                                   \\ \hline
\multirow{3}{*}{Food Delivery Apps} & \tcbox[on line, frame empty, boxsep=0pt,left=2pt,right=2pt,top=2pt,bottom=2pt,colback=LightYellow]{M5} Habitual Window Shopping                       & \multicolumn{1}{c}{\multirow{3}{*}{}}                                                                    \\ \cline{2-2}
                                    & \tcbox[on line, frame empty, boxsep=0pt,left=2pt,right=2pt,top=2pt,bottom=2pt,colback=LightYellow]{M6} Food Media-Triggered Use                        & \multicolumn{1}{c}{}                                                                                     \\ \cline{2-2}
                                    & \tcbox[on line, frame empty, boxsep=0pt,left=2pt,right=2pt,top=2pt,bottom=2pt,colback=LightYellow]{M7} Orders for Disordered Eating                   & \multicolumn{1}{c}{}                                                                                     \\ \Xhline{2\arrayrulewidth}
\end{tabular}
}
\caption{Motivations for digital food content consumption. The stated subcategories of digital food content are identified from open-ended interview responses. Note that these subcategories are not always distinctly separate from one another. Detailed descriptions of each subcategory are in our Supplementary Materials.} 
\label{table:motivation}
\end{table*}


\hr{Participants described a variety of motivations for engaging with digital food content. Table~\ref{table:motivation} outlines these motivations and their associated content subcategories. Overall, motivations for food media included both those linked to restrictive eating behaviors and those aimed at supporting recovery, whereas motivations for food delivery apps tended to align with disorder-exacerbating behaviors.}

We identified four motivations for seeking \textit{food media}. Many participants mentioned \textbf{\tcbox[on line, frame empty, boxsep=0pt,left=2pt,right=2pt,top=2pt,bottom=2pt,colback=LightYellow]{M1}expectations for vicarious satisfaction (N=7)}, \hr{often rooted in restrictive eating, a common eating disorder symptom characterized by severe limitation of food intake. They described intentionally avoiding actual food consumption, fearing weight gain, while watching others eat as a substitute. This mirrors the general population's use of eating broadcasts for vicarious satisfaction~\cite{anjani2020people}, but in our participants' case, the motivation was tied to dietary restriction rather than casual entertainment or sensory enjoyment.} 
For example, P21 said, ``\textit{If I overeat whenever I feel like eating, I'll gain weight and become obese. So, [instead of eating,] I watched it [food media] to feel vicarious satisfaction.}'' 

Some searched for videos filmed by people with EDs, such as binge eating broadcasts, with \textbf{\tcbox[on line, frame empty, boxsep=0pt,left=2pt,right=2pt,top=2pt,bottom=2pt,colback=LightYellow]{M2}expectations for comfort from shared experiences (N=4)}. Through those videos, they sought ``\textit{relief from shared experiences and feel solidarity}'' (P12). 
Every participant who remarked on expectations for relief from shared experiences noted that such media calm anxiety and stigma about having an ED. While previous research showed that the general population often watches eating broadcasts for a sense of social connectedness and attachment to the hosts~\cite{anjani2020people}, our finding reveals a different dimension: individuals with EDs are sometimes drawn to videos where the host engages in disordered eating. For them, this content offers not just a connection but a sense of comfort and validation in their shared struggles. This highlights a distinct psychological motivation closely tied to their experiences with their condition. 
\textbf{\tcbox[on line, frame empty, boxsep=0pt,left=2pt,right=2pt,top=2pt,bottom=2pt,colback=LightYellow]{M3}Seeking motivation for change (N=2)} was another motivator for watching such videos. To quote P2, ``\textit{[Looking at people with EDs binge eat food] made me realize that such behavior looks very bad to other people. This made me reflect on myself. ... Hence, I seek them to motivate myself to overcome my eating disorder.}'' 

Additionally, a few stated \textbf{\tcbox[on line, frame empty, boxsep=0pt,left=2pt,right=2pt,top=2pt,bottom=2pt,colback=LightYellow]{M4}expectations for relief by cooking along with the media (N=2)}. For instance, they engaged with cooking videos to ``\textit{alleviate the urge to binge eat and feel relief by cooking along with the videos}'' (P8). 
Interestingly, unlike the general population who often seek practical information such as recipes or nutritional details from cooking videos~\cite{wang2022tiktok}, some participants with EDs engage with this content for emotional relief and \hr{regulation of binge-eating urges,} 
highlighting a unique psychological motivation.

For the use of \textit{food delivery apps}, we identified three motivations. Notably, more than half of the participants who associated food delivery apps with their ED behaviors mentioned \textbf{\tcbox[on line, frame empty, boxsep=0pt,left=2pt,right=2pt,top=2pt,bottom=2pt,colback=LightYellow]{M5}habitual window shopping (N=6)} as a motivation for using food delivery apps. They ``\textit{browse food delivery apps when bored, but do not order anything}'' (P4). 
\hr{While seemingly casual, participants linked this habitual browsing to their disordered eating cycles, describing it as a precursor that often escalated into binge episodes. P17 described, ``\textit{I often check food delivery apps. When I feel [binge] urges, I open the app, put desired items in the cart, and if the order isn't accepted right away, I cancel it, thinking `I shouldn't binge eat it,' but then I end up ordering it again because I still want it. Even without urges, I browse out of boredom, see pictures, and feel triggered to want the food.}''} 

It is also notable that more than half of them mentioned \textbf{\tcbox[on line, frame empty, boxsep=0pt,left=2pt,right=2pt,top=2pt,bottom=2pt,colback=LightYellow]{M6}food media-triggered use (N=6)}. P6 described, ``\textit{After watching a video of somebody finishing a half gallon of mint chocolate ice cream, I ordered it (on a food delivery app). Often, I end up ordering and eating the same thing after watching it [eating broadcasts].}'' 
This suggests that combining different types of digital food content, specifically food media and food delivery apps, can together influence immediate behaviors among individuals with EDs. While food media on social media alone has been recognized as a powerful tool to encourage food product purchases~\cite{ventura2021socialfood}, its integration with food delivery apps appears to facilitate this effect. This dynamic interplay underscores the complex role of digital food content in shaping user behaviors, such as food choices and eating practices. 

Some participants said they use food delivery apps only to make \textbf{\tcbox[on line, frame empty, boxsep=0pt,left=2pt,right=2pt,top=2pt,bottom=2pt,colback=LightYellow]{M7}orders for disordered eating (N=4)}, but never for normal meals. They even said they often make multiple orders for disordered eating to get enough food to binge eat and purge. P7 said, ``\textit{When I binge eat, though I'm very full, I make multiple orders}''. 
\hr{This motivation reflects a direct and purposeful link to their ongoing eating disorder behaviors}

To summarize, \hr{motivations for food media ranged from those rooted in restrictive eating behaviors to recovery-supporting aims. This mix of restrictive-eating-driven and recovery-supporting motivations illustrates the ambivalence underlying food media engagement. On the other hand, motivations for food delivery apps were largely disorder-driven, that is, arising from and reinforcing existing patterns of disordered eating behaviors.}

\subsection{Digital Food Content Disorder-Exacerbating or Recovery-Supporting}
\label{res:dfc}

\begin{table*}[t]
\resizebox{\textwidth}{!}{
\begin{tabular}{lll}
\Xhline{2\arrayrulewidth}
\multicolumn{1}{c}{\textbf{Digital Food Content}} & \multicolumn{2}{c}{\textbf{How Being Associated with Disordered Eating}}                             \\ \hline\hline
\multirow{8}{*}{Food Media}               & \multirow{5}{*}{\tcbox[on line, frame empty, boxsep=0pt,left=2pt,right=2pt,top=2pt,bottom=2pt,colback=LightRed]{ED-Exacerbating Impact}}         & \tcbox[on line, frame empty, boxsep=0pt,left=2pt,right=2pt,top=2pt,bottom=2pt,colback=LightRed]{ED1} Stimulate Visually and Auditorily                    \\ \cline{3-3} 
                                                  &                                                     & \tcbox[on line, frame empty, boxsep=0pt,left=2pt,right=2pt,top=2pt,bottom=2pt,colback=LightRed]{ED2} Cannot Provide Vicarious Satisfaction          \\ \cline{3-3} 
                                                  &                                                     & \tcbox[on line, frame empty, boxsep=0pt,left=2pt,right=2pt,top=2pt,bottom=2pt,colback=LightRed]{ED3} Present Abnormally Large Amount of Food Consumption               \\ \cline{3-3} 
                                                  &                                                     & \tcbox[on line, frame empty, boxsep=0pt,left=2pt,right=2pt,top=2pt,bottom=2pt,colback=LightRed]{ED4} Trigger Overwhelming Thoughts about Food       \\ \cline{3-3} 
                                                  &                                                     & \tcbox[on line, frame empty, boxsep=0pt,left=2pt,right=2pt,top=2pt,bottom=2pt,colback=LightRed]{ED5} Normalize Eating Disorders from Watching Other Patients              \\ \cline{2-3} 
                                                  & \multirow{3}{*}{\tcbox[on line, frame empty, boxsep=0pt,left=2pt,right=2pt,top=2pt,bottom=2pt,colback=LightBlue]{Recovery-Supporting Impact}} & \tcbox[on line, frame empty, boxsep=0pt,left=2pt,right=2pt,top=2pt,bottom=2pt,colback=LightBlue]{REC1} Curb Urge to Binge Eating by Following Recipes \\ \cline{3-3} 
                                                  &                                                     & \tcbox[on line, frame empty, boxsep=0pt,left=2pt,right=2pt,top=2pt,bottom=2pt,colback=LightBlue]{REC2} Provide Relief through Shared Experiences              \\ \cline{3-3} 
                                                  &                                                     & \tcbox[on line, frame empty, boxsep=0pt,left=2pt,right=2pt,top=2pt,bottom=2pt,colback=LightBlue]{REC3} Promote Empowerment for Change                        \\ \hline
\multirow{2}{*}{Food Delivery Apps}                                & \multirow{2}{*}{ \tcbox[on line, frame empty, boxsep=0pt,left=2pt,right=2pt,top=2pt,bottom=2pt,colback=LightRed]{ED-Exacerbating Impact}}                           & \tcbox[on line, frame empty, boxsep=0pt,left=2pt,right=2pt,top=2pt,bottom=2pt,colback=LightRed]{ED1} Stimulate Visually                  \\ \cline{3-3}
& & \tcbox[on line, frame empty, boxsep=0pt,left=2pt,right=2pt,top=2pt,bottom=2pt,colback=LightRed]{ED6} Lower Barrier of Getting Food                  \\\Xhline{2\arrayrulewidth}
\end{tabular}
}
\caption{\hr{Recovery-supporting and disorder-exacerbating impacts of digital food content, along with associated contributing factors, illustrating their ambivalence.}
}
\label{table:dfc_lead_mitigate}
\end{table*}

\hr{The diverse motivations shaped how participants experienced digital food content, leading to both recovery‑supporting and \\ disorder‑exacerbating impacts. These contrasting effects often arose in relation to the same type of content, underscoring its ambivalent nature.} Table~\ref{table:dfc_lead_mitigate} outlines \hr{these impacts and the contributing factors that shaped them.} 
We identified five themes about how \textit{food media} contributes to eating disorder symptoms. Many participants discussed \textbf{\tcbox[on line, frame empty, boxsep=0pt,left=2pt,right=2pt,top=2pt,bottom=2pt,colback=LightRed]{ED1} visual and auditory stimuli (N=11)} incite their urge to binge eat. P10 said, ``\textit{I can't refrain myself from binge eating when I watch an eating broadcast with visual or auditory stimulation. The more my cravings are aroused [by watching it], the more and faster I eat.}'' 
P2 commented, ``\textit{the enticing display of food makes [the food media] more stimulating}.'' 
Some participants mentioned that the eating sound makes digital food media 
more invigorating. P17 said, ``\textit{it stimulates my appetite more when YouTubers eat food with vivid sounds.}'' 

Our participants also noted that, despite their expectations for vicarious satisfaction, watching food media \textbf{\tcbox[on line, frame empty, boxsep=0pt,left=2pt,right=2pt,top=2pt,bottom=2pt,colback=LightRed]{ED2} cannot provide them with vicarious satisfaction (N=6)} but instead increases the urge for food. P19 said, ``\textit{Looking back, I was falsely assuming that, oh, since the person in the video took my place and ate the food for me, I would be satisfied, which wasn't the case, and my urges piled up, leading me to binge eat. Afterward, I regretted my actions and vomited.}''  
Although a few studies claim that the general population gets vicarious satisfaction through watching eating broadcast~\cite{kircaburun2021psychology, anjani2020people}, our findings indicate that it does not always hold for people with EDs, for whom such content may instead lead to disordered eating behaviors.

Notably, every participant mentioned that the \textbf{\tcbox[on line, frame empty, boxsep=0pt,left=2pt,right=2pt,top=2pt,bottom=2pt,colback=LightRed]{ED3} abnormally large amount of food (N=22)} in food media has a negative impact, such as making them normalize binge eating. P6 said, ``\textit{People who eat a lot of food make me think, `It is okay to eat this much because there are other people who eat that much.}'~'' 
This finding extends previous research by showing that, while large food portions in eating broadcasts can lead the general population to normalize overconsumption~\cite{kircaburun2021problematic}, they can also make individuals with EDs normalize their disordered eating behaviors, such as binge eating.

Among all the participants, there was a unanimous assessment of eating broadcasts, indicating at least one adverse impact. However, participants had varying opinions on the effects of other food media types. For example, views on eating segments within non-food videos, such as vlogs, diverged. Some participants believed that eating segments were different from eating broadcasts and were not related to their ED symptoms. In contrast, others considered eating segments in non-food videos and eating broadcasts similar, attributing negative impacts to both.

Conflicting assessments also emerged regarding cooking posts and videos. Two participants who identified seeking relief as their motivation for watching digital food media (Section~\ref{res:motivation}), expressed a positive view of cooking posts and videos. They found these resources helpful in \textbf{\tcbox[on line, frame empty, boxsep=0pt,left=2pt,right=2pt,top=2pt,bottom=2pt,colback=LightBlue]{REC1} curbing the urge to binge eat by following the provided recipes (N=2)}. 
On the other hand, some participants experienced the opposite effect, as exposure to such media led them to ``\textit{continuously ruminate on the recipes and food}'' (P4), \textbf{\tcbox[on line, frame empty, boxsep=0pt,left=2pt,right=2pt,top=2pt,bottom=2pt,colback=LightRed]{ED4} triggering overwhelming thoughts about food (N=5)}.

For binge eating broadcasts, even the same individual experienced opposing impacts depending on the situation or circumstances.
They primarily fulfilled our participants' motivations, 
such as seeking \textbf{\tcbox[on line, frame empty, boxsep=0pt,left=2pt,right=2pt,top=2pt,bottom=2pt,colback=LightBlue]{REC2} relief through shared experiences (N=4)} and seeking \textbf{\tcbox[on line, frame empty, boxsep=0pt,left=2pt,right=2pt,top=2pt,bottom=2pt,colback=LightBlue]{REC3} empowerment for change (N=2)}. 
Moreover, they have made promises not to binge eat after watching binge eating broadcasts. P1 said, ``\textit{In the video, there are many instances where people regret and blame themselves for binge eating, so when I see that, I think, `I should eat in moderation, not like that.'~}''
In the meantime, P6 stated that she sometimes felt that she does ``\textit{not have to treat eating disorder, as many people do.}'' This \textbf{\tcbox[on line, frame empty, boxsep=0pt,left=2pt,right=2pt,top=2pt,bottom=2pt,colback=LightRed]{ED5} normalization of EDs from watching other patients (N=2)} could impede the recovery process for individuals with EDs.

Regarding how \textit{food delivery app} use contributes to eating disorders, we identified two themes about how it assists disordered eating. 
One common impact between food media and food delivery apps was \textbf{\tcbox[on line, frame empty, boxsep=0pt,left=2pt,right=2pt,top=2pt,bottom=2pt,colback=LightRed]{ED1} stimulating visually (N=2)}. P10 noted, ``\textit{With habitual use of food delivery apps, the enticing display of food stimulates my cravings.}'' 
%
In addition, many participants mentioned the \textbf{\tcbox[on line, frame empty, boxsep=0pt,left=2pt,right=2pt,top=2pt,bottom=2pt,colback=LightRed]{ED6} lower barrier of getting food (N=11)} on delivery apps compared with buying them in-person as a significant factor leading to disordered eating. P11 said, ``\textit{Food delivery apps enabled us to conveniently obtain food. People bring food to my door only with a few clicks. So I don't need to be self-conscious of how I look to other people. I can finish the (binge eating) process as if nothing happened. This is really bad.}'' 

In conclusion, \hr{digital food content had both recovery-supporting and disorder-exacerbating impacts on participants with ED, a coexistence we frame as \emph{impact ambivalence}. This ambivalence manifested differently depending on the individuals, circumstances, and subcategories.}

\subsection{Efforts to Avoid Digital Food Content Leading to Disordered Eating and Relapse}
\label{res:efforts}

\begin{table*}[t]
\resizebox{.65\textwidth}{!}{
\begin{tabular}{ll}
\Xhline{2\arrayrulewidth}
\textbf{Digital Food Content}       & \tcbox[on line, frame empty, boxsep=0pt,left=2pt,right=2pt,top=2pt,bottom=2pt,colback=LightViolet]{\textbf{Avoidance Practices}}                  \\ \hline\hline
Food Media                          & \tcbox[on line, frame empty, boxsep=0pt,left=2pt,right=2pt,top=2pt,bottom=2pt,colback=LightViolet]{A1}Influence the Algorithm to Reduce Suggestions \\ \hline
\multirow{3}{*}{Food Delivery Apps} & \tcbox[on line, frame empty, boxsep=0pt,left=2pt,right=2pt,top=2pt,bottom=2pt,colback=LightViolet]{A2}Uninstall Food Delivery Apps                   \\ \cline{2-2} 
                                    & \tcbox[on line, frame empty, boxsep=0pt,left=2pt,right=2pt,top=2pt,bottom=2pt,colback=LightViolet]{A3}Delete Accounts                               \\ \cline{2-2} 
                                    & \tcbox[on line, frame empty, boxsep=0pt,left=2pt,right=2pt,top=2pt,bottom=2pt,colback=LightViolet]{A4}Use Shared Accounts                            \\ \Xhline{2\arrayrulewidth}
\end{tabular}
}
\caption{Practices for avoiding digital food content leading to disordered eating.}
\label{tab:avoidance_practice}
\end{table*}

\hr{Aware of the potential harms from digital food content, many participants attempted to avoid it. However, these efforts were often short‑lived, with relapses driven either by a loss of self‑control or the anticipation of recovery‑supporting effects they had previously experienced, such as vicarious satisfaction. We examine the avoidance strategies participants used, the challenges in sustaining them, and the recurring cycle of disengagement and return. Table~\ref{tab:avoidance_practice} summarizes these avoidance practices. }

\hr{For \textit{food media}, eleven participants reported deliberate efforts to reduce or remove exposure.} 
Most mentioned that they tried to \textbf{\tcbox[on line, frame empty, boxsep=0pt,left=2pt,right=2pt,top=2pt,bottom=2pt,colback=LightViolet]{A1}influence the algorithm to reduce suggestions~(N=8)} of food videos by unsubscribing from eating broadcast channels and/or deliberately searching for non-food content. 
They utilized the `Not interested' and `Don't recommend channel' buttons on YouTube and the `Not interested' button on Instagram to remove food media from their personalized algorithm. P11 said, ``\textit{There are different measures you can take to make the (YouTube) algorithm not recommend certain videos. So, I tried to use them to remove recommendations for food-related videos.}'' 
In addition, P13 ``\textit{deliberately searched for random keywords to change recommendations.}''

Every participant who tried to modify their suggestion algorithm noted that such efforts often failed, \hr{as they eventually returned to actively searching for food-related videos, sometimes to satisfy urges and seek vicarious satisfaction, such as watching eating broadcasts to curb food cravings. }
The algorithm quickly responds to such behaviors. P2 said, ``\textit{\hr{Even after putting effort into removing food videos from my recommended feed, I eventually gave in to my urges and searched for them again. They quickly showed up on my feed once more. For example, when I felt hungry, I intentionally returned to such videos to ease the hunger, but that became the starting point, and I ended up eating a lot even when I was already full.}}'' 
\hr{Some participants described returning to food media not only because of sudden urges but also to recapture the comfort or satisfaction they had previously felt, such as vicarious satisfaction from watching eating broadcasts without actually consuming food.}

Eleven participants reported that they tried to regulate \emph{food delivery app} use. 
All participants mentioned \textbf{\tcbox[on line, frame empty, boxsep=0pt,left=2pt,right=2pt,top=2pt,bottom=2pt,colback=LightViolet]{A2}uninstalling food delivery apps (N=11)} and some stated \textbf{\tcbox[on line, frame empty, boxsep=0pt,left=2pt,right=2pt,top=2pt,bottom=2pt,colback=LightViolet]{A3}deleting their accounts (N=3)}. 
However, everyone described such efforts often prove ineffective as they tend to reinstall the apps whenever they lose self-control, as reinstalling or creating new accounts is trivial. 
To quote P9, ``\textit{I repeatedly uninstall food delivery apps trying not to binge eat, but reinstall it whenever I want to binge eat.}'' 

Two participants mentioned \textbf{\tcbox[on line, frame empty, boxsep=0pt,left=2pt,right=2pt,top=2pt,bottom=2pt,colback=LightViolet]{A4} using shared accounts~(N=2)} as a strategy to have an extra layer of vigilance in managing their engagement with food delivery apps. They explained that, by using shared accounts linked with family members, they introduced an element of accountability even when they were alone. For instance, P4 shared, ``\textit{If my parents are at work and I am home alone, I might order and binge eat something, and destroy the evidence, and they wouldn't know what I did. So, there is an option for a family account. Yes, I have set that up.}'' 
This approach allowed them to create a safeguard against impulsive app usage, as their actions would be visible to others, potentially discouraging disordered eating. 
Note, however, that not every food delivery app offers the option of shared accounts.

\hr{In summary, our participants engaged with digital food content for diverse reasons and across multiple subcategories. While this engagement often led to ED-exacerbating impacts, some participants also turned to certain types of food media, such as cooking videos or binge eating broadcasts, as a way to alleviate their symptoms or support recovery. Across both food media consumption and food delivery app use, we observed a recurring pattern in which participants attempted to self-regulate their usage to prevent adverse impacts, yet frequently relapsed. These relapses were sometimes driven by sudden urges, but also by the anticipation of previously experienced benefits, resulting in an ambivalence‑driven cycle where the same content’s recovery-supporting and harmful aspects alternately pulled participants toward and away from engagement.}

\section{Discussion}
\label{sec:discussion}

Our study highlighted the conflicting impacts of digital food content on individuals with EDs, perpetuating the cycle in which they regulate their consumption to avoid ED-exacerbating impacts, yet eventually return when losing self-control or with the expectation of supportive effects on their condition. In this section, we discuss the process underlying their content consumption behavior based on dual systems theory~\cite{norman1986attention}, introducing the concept of \emph{ambivalence} (Section~\ref{sec:discussion:ambivalence}).

\subsection{Impact Ambivalence: Cognitive Mechanism of Digital Food Content Consumption} 
\label{sec:discussion:ambivalence}


\begin{figure*}
  \begin{subfigure}{0.4\textwidth}
    \includegraphics[width=\linewidth]{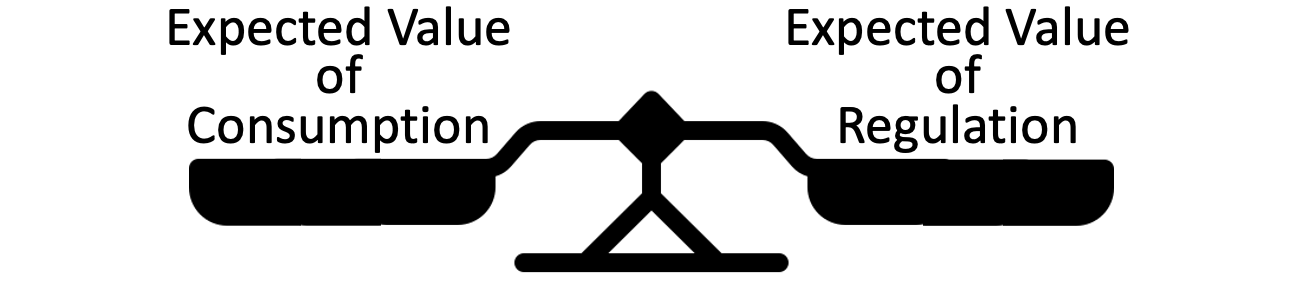}
    \caption{Evaluation of consuming vs. not consuming digital food content, where different actions lead to distinct, non-conflicting outcomes.} \label{fig:3a}
  \end{subfigure}%
  \hspace*{1.3cm}   
  \begin{subfigure}{0.44\textwidth}
    \includegraphics[width=\linewidth]{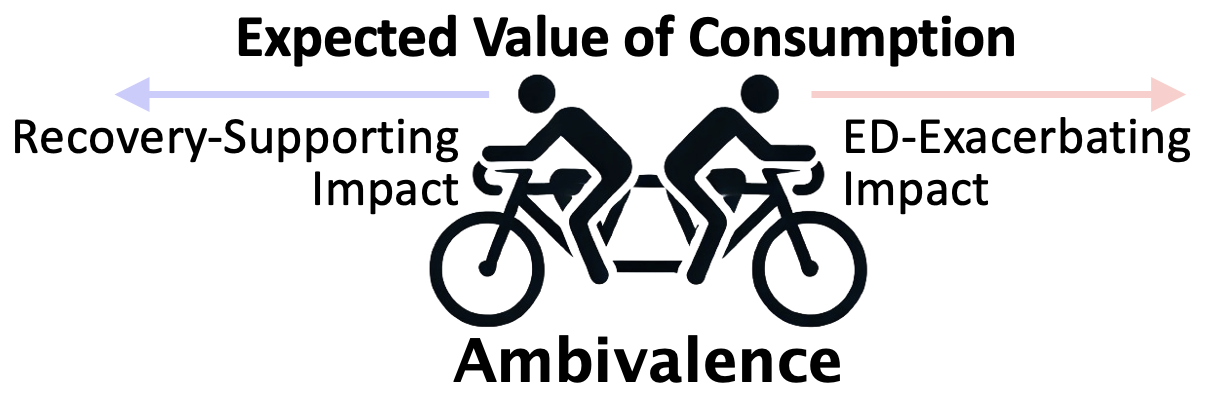}
    \caption{Ambivalence in assessing the expected value of consumption, where the same action (engaging in digital food content) may lead to conflicting outcomes, such as recovery-supporting and/or ED-exacerbating impacts.} \label{fig:3b}
  \end{subfigure}%
\vspace{-0.3cm}
\caption{Conceptual illustrations of expected value assessments of engaging with digital food content.}
\Description{Figure 2 conceptually illustrates expected value assessments associated with engaging with digital food content. (a) depicts a scale weighing the expected value of content consumption against that of self-regulation. (b) illustrates a bicycle with two people pedaling in opposite directions, symbolizing the conflicting impacts of content consumption.}
\end{figure*}

A recent study framed the cognitive mechanisms of digital food content consumption behavior by individuals with EDs using the extended dual systems model of self-regulation~\cite{choi2024foodcensor}. Dual systems model~\cite{norman1986attention} posits two distinct cognitive decision-making systems: system~1, which is fast, intuitive, and automatic responses, and system~2, which is slow, deliberate, and analytical thinking. System~1 handles routine or immediate decisions, while system~2 engages in more effortful, reflective thinking for complex or high-stakes decisions, ensuring rationality and accuracy but requiring more cognitive resources. 
These systems interact dynamically, with system~1 providing quick responses (e.g.,~feeling an impulse to watch a food video when encountering it), which system~2 evaluates (e.g.,~assessing whether watching the food video aligns with personal ED-recovery goals), shaping how we process information and make decisions. When framing the cognitive mechanisms of digital food content consumption, prior work has largely emphasized the adverse effects of digital food content, assuming that individuals use system 2 primarily to regulate consumption in order to minimize harm ~\cite{choi2024foodcensor}.

However, our study revealed that individuals with EDs sometimes deliberately engage with digital food content to support recovery, such as by drawing motivation from various types of food media. This observation challenges the assumption in prior work that system 2's role is solely harm-avoidant. 
Instead, our findings suggest that system~2 also incorporates the potential positive impacts of engaging with such content. In other words, when assessing the expected value of their digital food content consumption behavior, individuals may navigate two layers of decision-making: first, comparing the expected outcomes of consuming versus not consuming digital food content as suggested in the prior work (Figure~\ref{fig:3a}); and second, encountering conflicting expected values, recovery-supporting benefits and the risk of adverse effects, within the act of consumption itself (Figure~\ref{fig:3b}).


\hr{We describe this conflict as \emph{impact ambivalence}, extending the notion of attitudinal ambivalence, the coexistence of positive and negative evaluations toward the same behavior~\cite{thompson2014let, van2009agony, schneider2017mixed, van2015abc}. Impact ambivalence extends this idea to the level of behavioral consequences, where a single act of engagement can generate both supportive and harmful consequences. In our context, digital food content engagement is not just judged ambivalently but experienced ambivalently, producing both recovery-supporting and disorder-exacerbating outcomes. This makes engagement inherently tense, as individuals feel simultaneously drawn to and resistant to consumption, complicating System 2 evaluations with effects that are difficult to reconcile. Because these consequences are often unpredictable, individuals cannot reliably anticipate whether engagement will support recovery, exacerbate symptoms, or do both. This underscores the need for interventions that operate during digital food content engagement,} mitigating negative impacts while allowing potential positive effects to remain. In the following, we present design implications to foster healthier engagement with digital food content while addressing its ambivalent impacts.

\subsection{Promoting Informed and Healthier Content Engagement}
\label{sec:discussion:informed}

\begin{figure*}[t]
    \centering
    \includegraphics[width=\textwidth]{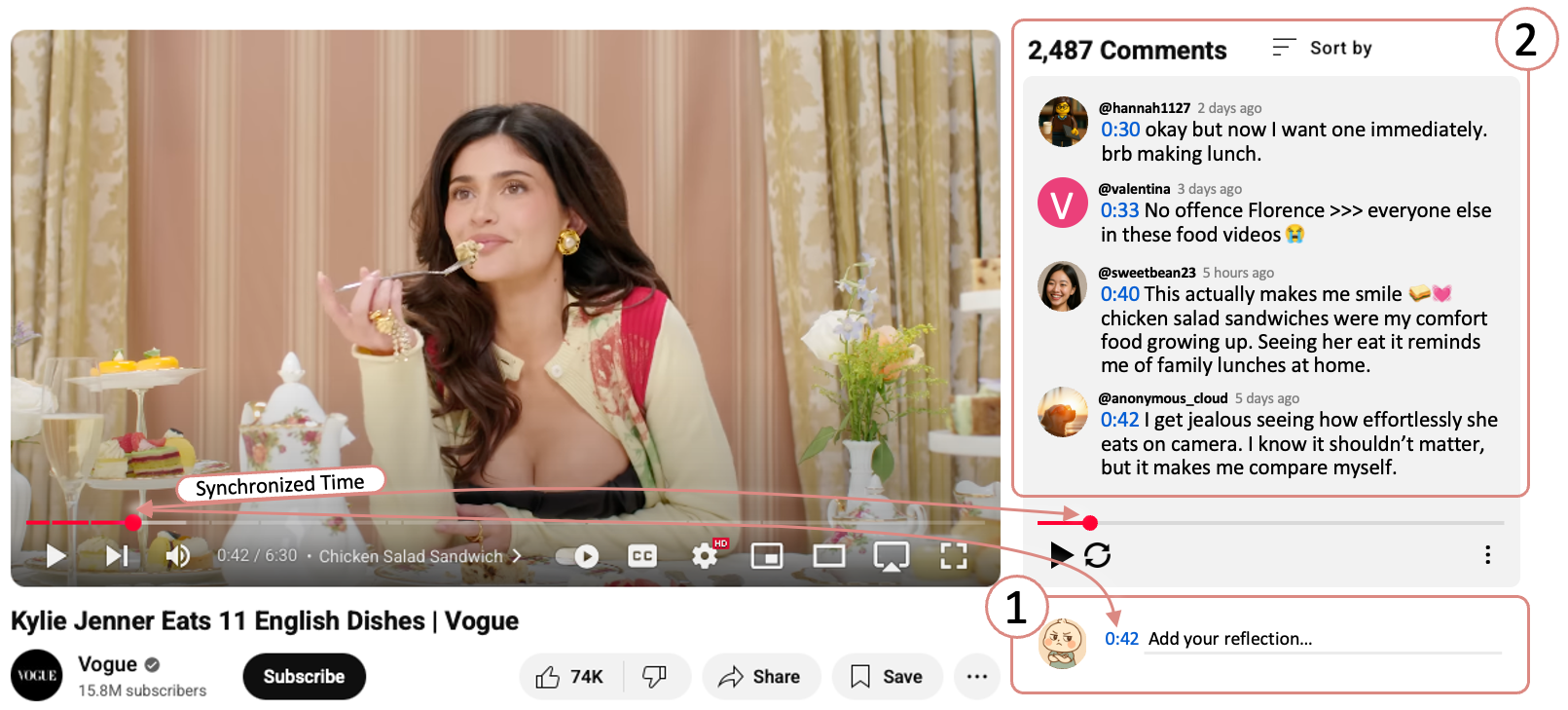}
    \vspace{-0.6cm}
    \caption{Example wireframes portraying design implications for reflective and collective feedback to surface media's various impacts. \circled{1}~Foster users to reflect on and share their feelings and thoughts while watching. \circled{2}~Provide visibility into others' in-the-moment reflections, surfacing how content may be experienced differently across viewers.}
    \Description{Figure 3 is an example wireframe mock-up of a video player interface showing synchronized reflective and collective feedback features. The left side displays a paused food video at timestamp 0:42. Below the video timeline, a synchronized marker links the current time with a reflection entry box, prompting the viewer to add a comment. On the right side, a live comment panel shows time-stamped reflections from other viewers at nearby moments, including emotional responses, personal memories, and comparison concerns. Numbered callouts highlight two design implications: (1) enabling users to write reflections while watching and (2) allowing visibility into others’ in-the-moment reflections.}
    \label{fig:feedback_media}
\end{figure*}

Research on decision-making has emphasized the importance of \emph{informed} decision-making~\cite{stacey2017decision, lindrup2021one}. Access to accurate, complete, and contextually relevant information helps individuals evaluate options and align decisions with their goals, values, or needs. Specifically, in dual systems theory, information facilitates system~2's deliberate and analytical thinking to match their decisions with their goals. 
We found individuals with EDs have engaged with digital food content either by expecting positive impacts or through unconscious, habitual consumption. Within the framework of dual systems theory, this indicates that engagement decisions are often driven by system~2's expected value assessments, focusing mainly on positive outcomes, or by system~1's automatic processes. Meanwhile, our study stresses that engaging with digital food content sometimes carries ED-exacerbating impacts misaligned with individuals' expected recovery-supporting benefits, prompting our participants to spontaneously self-regulate their consumption behaviors. In this regard, we suggest that it is crucial to provide individuals with information that helps them recognize and evaluate the potential negative impacts of digital food content alongside the positive ones in their engagement decision-making process. By incorporating this information into system~2's expected value assessment, individuals could make balanced choices.

\subsubsection{Reflective and Collective Feedback to Uncover Media’s Ambivalent Impacts} 

Our findings reveal that individuals with EDs engage with and respond to digital food content differently from the general population. While media platforms often implement content warnings to protect vulnerable users (e.g.,~content warnings on violent content for children)~\cite{charles2022typology, x2024your, netflix2021photosensitivity}, these warnings are typically designed for broad audiences and struggle to capture the nuanced needs of diverse populations~\cite{haimson2020trans}. As a result, less visible harms, such as the exacerbating impacts of digital food content on individuals with ED, are often overlooked.

To address these limitations, future research should investigate alternative deliberate approaches that could surface both overlooked harms and benefits. One promising direction is enabling viewers to articulate and share their emotional and cognitive responses while watching videos~(Figure~\ref{fig:feedback_media}~\circled{1}). This aligns with evidence that expressive writing can facilitate reflection and meaning-making, and that articulating reactions in situ can heighten self-awareness~\cite{smyth1998written}. Building on this, future research could also investigate interventions that make such reflections visible across viewers (Figure~\ref{fig:feedback_media}~\circled{2}), offering not only individual insight but also transparency into how the same media can be experienced differently by others.

When aggregated across multiple viewers, these responses could provide a form of \emph{social translucency}, surfacing how different segments and features of content are experienced. The aggregated view may highlight collective patterns---where reactions tend to converge---as well as points of divergence across subgroups, revealing how the same content may be interpreted differently depending on viewers' characteristics or vulnerabilities. This collective layer could help media platforms and creators identify where ambivalent impacts occur, for example, in our context, moments that simultaneously elicit recovery-supporting benefits and disorder-exacerbating risks. Further research is needed to explore how such insights might be leveraged to detect critical ambivalent impacts and explore ways of mediating them toward more positive outcomes. Rather than relying solely on one-size-fits-all warnings, viewer-informed signals may open pathways toward more personalized support for informed engagement. 
We note, however, that such interventions could introduce unintentional risks, and future work should therefore investigate how reflective and collective feedback can be responsibly incorporated to support informed media engagement without creating new challenges.

\subsubsection{Prompts to Support Mindful Usage of Food Delivery Apps}

We identified a recurring pattern of uninstalling and reinstalling food delivery apps among individuals with EDs to avoid unhealthy usage of these apps. To foster healthier engagement, we suggest interventions to promote mindful food delivery app use. 
For instance, a recent study demonstrated the effectiveness of reflective prompts in encouraging mindful engagement with food-related videos on YouTube for individuals with EDs~\cite{choi2024foodcensor}. In that study, participants were shown short prompts, such as ``Why do you want to watch this video now?'', which encouraged them to pause, evaluate their motivations, and sometimes disengage from potentially triggering content. These prompts did not block access but created a small reflective moment that supported more deliberate media choices. In the context of food delivery apps, a similar in-the-moment intervention could be triggered upon app entry, for example, asking, ``\textit{Will this order support your recovery goals today?}''. Such prompts could support mindful app usage~\cite{xu2022typeout}. A potential advantage is that these lightweight interventions respect autonomy and allow users to continue if they choose, while still nudging reflection. However, drawbacks include the risk of alert fatigue if prompts appear too frequently, or the possibility that prompts may feel judgmental and increase guilt, which could undermine recovery.


\subsubsection{Opportune Moments to Prevent Disordered Eating Triggered by Digital Food Content}
\label{sec:discussion:opportune}

The impact ambivalence of digital food content makes it difficult for our participants to assess the outcomes of engaging with such content. This unpredictability poses significant challenges for intervention design, particularly in determining when and how to intervene. For instance, interventions that aim to regulate engagement with digital food content to prevent its negative impacts~\cite{choi2024foodcensor} might inadvertently block opportunities for users to gain recovery-supporting benefits. In this regard, further intervention designs for healthy engagement with digital food content should account for the complex mechanisms underlying its conflicting impacts. 
To guide the design of effective interventions, we suggest that researchers explore how users interact with different types of digital food content. These interactions are influenced by various motivations and impacts, and understanding them can help create strategies to prevent disordered eating behaviors.

Recognizing the uncertainty of impacts, our study identified a few moments that could offer a second chance to steer users away from the impacts that exacerbate EDs subsequent to engaging with digital food content. Some participants indicated prolonged browsing or frequent engagement with food delivery apps often precedes binge eating episodes, implying opportune moments for situated intervention. 
In addition, our participants reported that food media sometimes triggered their use of food delivery apps, driven by the urge to eat the foods featured due to the media's visual and auditory stimuli. This sequence of engaging with food media and food delivery apps could imply their ED-exacerbating impacts, such as visual and auditory stimuli of food media and the low barrier of getting food from food delivery apps. These usage patterns detail the practices associated with ED-exacerbating impacts of digital food content, suggesting critical moments for timely interventions to mitigate its adverse effects. 

Another opportune moment we propose to leverage is the waiting period between ordering and food delivery on food delivery apps. Previous research has explored digital interventions such as showing videos that teach mindful eating or promote healthy body images just before meals to encourage healthy eating behaviors~\cite{allirot2018effects, zepeda2008think}. However, these interventions often require proactive user engagement, such as manually opening and watching the videos on their devices. By detecting associated app features or screens (e.g.,~food waiting screens), such interventions could be passively triggered and seamlessly integrated into the user experience. 
For instance, an intervention could prompt users with short, guided questions, such as reflecting on their feelings or hunger levels to promote mindful eating~\cite{luo2019co}, while waiting for delivery. Such strategies could help foster self-reflection on food ordering, reduce impulsivity, and disrupt planned binge eating.

\subsection{Mindful Curation of Digital Food Content for Dietary Well-being}
\label{sec:discussion:curation}

In the field of human-food technology interaction, there has been a growing emergence of research focused on promoting healthy eating~\cite{khot2022understanding, covaci2023no}. Simultaneously, a substantial body of research has been dedicated to improving the eating experience through enhancing the realism or stimuli of virtual food~\cite{velasco2016multisensory, anjani2020people}. Recent studies have also investigated how tech-mediated auditory stimuli, such as different music playing or manipulating eating sounds (e.g.,~amplifying and delaying)~\cite{peng2021sounds, kleinberger2023auditory}, affect users' eating practices. Another research demonstrated that subtly adjusting the playback speed of food videos during eating could promote slower eating and more controlled intake~\cite{chen2025vifeed}. Understanding the impact of various stimuli related to food, as delivered through various technologies, on eating practices is essential. Coupled with our study, we believe this understanding could create a more health-conscious and enjoyable experience in food multimedia design and curation, ultimately enhancing users' experiences.

Our findings stress the importance of carefully curating food multimedia content that is increasingly captivating, engaging, and often exacerbating disordered eating~\cite{kinkel2022food}. Mindful digital food content curation is particularly important when promoting healthy eating among populations highly susceptible to the influential nature of digital content, such as individuals with EDs~\cite{choi2024foodcensor}. 
Contrary to the prevailing content recommendation strategy, which reinforces users' exposure to similar content by relying on their recent viewing history to attract users~\cite{chaney2018algorithmic, youtubeSuggestionAlgorithm}, a more mindful curation approach is needed. 

For instance, multimedia platform designers could consider adopting a counterbalancing approach to prevent users from becoming trapped in the `filter bubble,' where they are repeatedly exposed to the same type of content~\cite{nguyen2014exploring}. For example, when users have heavily engaged with food media, the platform could strategically introduce content on different topics. This intentional diversification of content could help mitigate unconscious engagement with triggering food media for individuals with EDs by reducing their passive exposure to food-related media. Recent research also demonstrated that reducing exposure to food-related videos on YouTube can alleviate their negative impacts on individuals with EDs, leading to fewer obsessive thoughts about foods and overall improvement in quality of life~\cite{choi2024foodcensor}.

However, implementing such a counterbalancing strategy requires carefully considering the overall user experience. If not thoughtfully designed, this approach could disrupt the flow of content the users find engaging, leading to frustration or reduced platform satisfaction~\cite{wei2023measuring}. Therefore, future work should explore balancing content curation that breaks the filter bubble, supporting users' well-being while maintaining the recommended content attractive and engaging. Additionally, future research could explore strategies for providing a counterbalance to users and assessing its impact on both user engagement and health outcomes.

\subsection{Limitations}
Our study predominantly included participants residing in South Korea, with only two individuals from North America and Europe. While food media and food delivery apps have achieved global popularity~\cite{anjani2020people, onlineFoodDelivery}, the limited demographic diversity of our participants means that our findings may not be fully generalizable across different cultural and geographic contexts. 

\section{Conclusion}

We examined how individuals with eating disorders perceive and interact with digital food content. Our study offers insights into the complex dynamics between individuals with eating disorders and digital food content, uncovering its ambivalent roles as a recovery-supporting tool and a trigger for disordered eating behaviors. We demonstrated how this \emph{impact ambivalence} creates significant tension, perpetuating behavioral patterns characterized by cycles of self-regulated content engagement and relapse. Our design implications and proposed strategies for promoting mindful engagement with digital food content, and \hr{the potential opportune moments we identified to mediate its negative} aftereffects, \hr{together provide} a clear foundation for HCI designers to develop more personalized and supportive digital environments.  
By addressing individuals with eating disorders' unique needs, HCI researchers and practitioners could contribute to shaping healthy and empowering digital spaces. Our research contributes to the HCI discourse by examining the potential for more responsible and health-conscious design practices in digital food content.




\newpage
\bibliographystyle{ACM-Reference-Format}
\bibliography{main}


\end{document}